\documentclass[
reprint,
superscriptaddress,
aps,
prd,
amsmath,
amssymb,
longbibliography,
nofootinbib,
showkeys
]{revtex4-2}

\usepackage[T1]{fontenc}
\usepackage[utf8]{inputenc}
\usepackage[english]{babel}

\usepackage{amsfonts}
\usepackage{bm}
\usepackage{mathtools}

\usepackage{graphicx}
\usepackage[caption=false]{subfig}
\usepackage{tikz}

\usepackage{booktabs}
\usepackage{array}
\usepackage{multirow}

\usepackage{siunitx}

\usepackage{xcolor}

\usepackage[
colorlinks=true,
linkcolor=blue,
citecolor=blue,
urlcolor=blue
]{hyperref}

\usepackage{orcidlink}

\newcommand{\rev}[1]{\textcolor{blue}{#1}}

\providecommand{\doi}[1]{%
  \href{https://doi.org/#1}{doi:#1}%
}

\begin{document}


\title{
Conformal Geometry and Regularization of Disclinations
by a Cosmological Constant in (2+1) Dimensions
}


\author{A.~M.~de~M.~Carvalho\,\orcidlink{0009-0006-3540-0364}}
\email{alexandre@fis.ufal.br}
\affiliation{
Instituto de Física,
Universidade Federal de Alagoas,
57072-970 Maceió, AL, Brazil
}

\author{C.~Furtado\,\orcidlink{0000-0002-3455-4285}}
\email{furtado@fisica.ufpb.br}
\affiliation{
Departamento de Física,
Universidade Federal da Paraíba,
58051-970 João Pessoa, PB, Brazil
}


\begin{abstract}
We investigate the effect of a cosmological constant $\Lambda$
on the geometry generated by a two-dimensional disclination within a conformal metric framework. For $\Lambda>0$, we obtain an exact
analytic solution of the Liouville-type equation, providing a natural regularization of the defect while preserving its local conical structure. The resulting geometry possesses constant positive scalar curvature, $R=3\Lambda$. For $\Lambda<0$, no real closed-form solution was found within the present approach; the corresponding numerical solutions approach an asymptotically hyperbolic geometry with $R\to 3\Lambda<0$. The analysis shows that the cosmological constant determines the curvature scale and asymptotic behavior of the regularized geometry, while the disclination strength $\alpha$ remains encoded in its local topology.
\end{abstract}




\maketitle


\section{Introduction}
\label{intro}

Analog models of gravity provide powerful frameworks for investigating emergent spacetime structures within condensed matter systems. The pioneering work of Unruh~\cite{Unruh1981} demonstrated that perturbations in moving fluids propagate as if in a curved spacetime, thereby establishing the foundation of analog gravity. This idea was further developed in Refs.~\cite{Visser1998,Barcelo2001,Barcelo2005}, where it was shown that Bose–Einstein condensates (BECs) and other condensed matter systems can give rise to effective geometries exhibiting key features such as horizons, ergoregions, and Hawking-like radiation.

A particularly intriguing aspect of analog gravity lies in the emergence of vacuum energy contributions within effective gravitational descriptions. In the context of superfluid \( ^3 \)He-A, Volovik~\cite{Volovik1986,Volovik1998,Volovik2003,JannesVolovik2012,Finazzi2012} demonstrated that an effective cosmological constant can arise from the energy difference between the equilibrium (true vacuum) and perturbed states of the system. Within this framework, the analog cosmological term captures how the ground state responds to geometric or topological deformations, functioning as a measure of vacuum rigidity. This interpretation echoes broader perspectives such as that of Carroll~\cite{Carroll2001}, where the cosmological constant is viewed as a manifestation of vacuum energy density and its associated pressure in the context of general relativity.

Topological defects offer a natural setting for studying geometric perturbations in physical systems. Disclinations, in particular—line-like defects associated with angular mismatch—arise in a variety of contexts, including liquid crystals, grain boundary networks in polycrystalline solids, and two-dimensional materials~\cite{Kleinert1989,Kleman2008,Fumeron2022,Fumeron2023}. These defects locally break rotational symmetry and concentrate curvature along their axis. Their idealized geometric representation as conical singularities provides a powerful framework for exploring the interplay between topology, curvature, and material response.

While conical metrics successfully capture the local structure of disclinations, 
they leave open a central question: how does the inclusion of a cosmological constant 
$\Lambda$ modify the geometry and the physical content of such defects? More specifically, 
can $\Lambda$ regularize the asymptotic behavior of the metric and smooth out the singular 
core in a way that better reflects realistic condensed-matter systems or analog spacetimes?

While previous works have treated disclinations either without vacuum contributions 
or through cut-off dependent schemes, here we show that a cosmological constant 
provides a natural and analytic mechanism for regularization.

In this work, we analyze the geometry induced by a planar disclination in the presence 
of a cosmological constant. Within a conformal metric framework, we derive and exactly solve 
the modified field equation for the conformal factor associated with a point-like disclination source. 

The resulting geometry interpolates between a classical conical core and a constant-curvature background, with the cosmological constant controlling both the regularization of the defect and the large-distance structure of the geometry. 

To the best of our knowledge, this is the first study to address disclination geometry 
with a nonzero cosmological constant in a fully regularized conformal setting. The model 
establishes a unified framework for interpreting curvature and topological charge in the 
presence of vacuum energy--like contributions, opening new perspectives for analog gravity, 
emergent curved phases, and geometric approaches to defect theory in condensed-matter systems.

The remainder of this paper is organized as follows. Section~\ref{topological} reviews the geometric theory of topological defects in two-dimensional media, with emphasis on disclinations and their conical description. Section~\ref{geometric} introduces the conformal metric formalism used to model the curvature generated by such defects. Section~\ref{motivation} motivates the inclusion of a cosmological constant and derives the resulting nonlinear field equation. Section~\ref{sec:regularization} presents the exact $\Lambda$-regularized solution and analyzes both its near-core and asymptotic regimes. Section~\ref{sec:scalar} computes the scalar curvature and highlights the role of the sign of $\Lambda$ in the global structure. Section~\ref{conclusion} summarizes the results and outlines perspectives for future work.


\section{Topological Defects in Two-Dimensional Media}
\label{topological}

Many physical systems—including nematic liquid crystals, two-dimensional materials, and models of analog gravity—exhibit line-like topological defects that arise from spontaneous symmetry breaking and the nontrivial topology of the order parameter field. These defects correspond to singular configurations that cannot be removed by smooth deformations, giving rise to intrinsic geometric and topological structures in the material or effective spacetime~\cite{Kleman2008,Addou2022,Moraes2000}.

Two fundamental types of line defects are typically encountered: disclinations, associated with broken rotational symmetry and curvature concentration, and dislocations, related to broken translational symmetry and torsional effects. In the geometric theory of defects~\cite{katanaev1992}, disclinations are modeled by conical geometries with angular deficits or excesses, whereas dislocations are described through torsional singularities.

Experimental and computational studies in colloidal crystals and atomic lattices have demonstrated that such defects are not merely static singularities. Under suitable conditions, disclinations and dislocations can emerge spontaneously, move through the medium, interact elastically, and even annihilate when oppositely charged defects meet. In systems with thermally activated dynamics, bound pairs of defects may also separate, leading to the proliferation of free topological excitations~\cite{Chen2008,Addou2022,Rafayelyan2024}.

Recent developments have generalized this geometric description to account for continuous distributions of defects in two-dimensional materials such as graphene. In these systems, curvature and torsion emerge not from isolated singularities, but from extended defect structures and grain boundaries~\cite{Addou2022,s41524-022-00871-y,carvalho2025geometric}. These advances highlight the importance of incorporating smooth geometric deformations and effective curvature responses into the modeling of realistic defect configurations.

In this work, we focus on curvature–inducing line defects (disclinations) and investigate how their geometric description changes when a cosmological-constant term is added to the two-dimensional conformal metric. We solve exactly, for $\Lambda>0$, the Liouville-type equation with a point source and show that $\Lambda$ acts as an infrared regulator: it smooths the core while preserving the near-core conical charge $\alpha$, and introduces a crossover scale $r_c\sim a^{-1/\alpha}$ that separates the defect-dominated region from the $\Lambda$-dominated regime.
For $\Lambda<0$, numerical solutions reveal an asymptotically hyperbolic geometry with $R\to 3\Lambda<0$. 
This provides a cosmological regularization of the classical conical model,
motivated by analog-gravity ideas and effective geometries in condensed-matter settings.

The present study is theoretical. Possible observables associated with
the regularized geometry, including LDOS variations and holonomy phases,
are not addressed here and are left for future investigation. A direct
experimental realization of the $\Lambda$-regularized defect likewise
remains an open direction.

\section{Geometric Theory of Topological Defects}
\label{geometric}

The geometric theory of topological defects, developed by Katanaev and Volovich~\cite{katanaev1992,Katanaev2005}, provides a unified framework for describing curvature and torsion induced by line-like defects in continuous media. In this approach, the medium is treated as a Riemann–Cartan manifold equipped with a metric \( g_{\mu\nu} \), a curvature tensor \( R^\mu{}_{\nu\alpha\beta} \), and a torsion tensor \( T^\mu{}_{\alpha\beta} \). Disclinations, which result from broken rotational symmetry, are associated with curvature, while dislocations, arising from broken translational symmetry, are modeled as sources of torsion. When torsion vanishes, the geometry reduces to a purely Riemannian structure, appropriate for curvature-based defects such as disclinations.

For two-dimensional systems with cylindrical symmetry and translational invariance along the defect axis (\(z\)), the metric can be expressed in conformal form\cite{Carvalho2013}:
\begin{equation}
    ds^2 = e^{2\Omega(r)} (dr^2 + r^2 d\theta^2) + dz^2,
    \label{eq:metric_conformal}
\end{equation}
where the conformal factor \( \Omega(r) \) encodes the radial dependence of the curvature generated by the defect.
Because the geometry is translationally invariant along the defect axis $z$,
the problem reduces to the two-dimensional transverse section $(r,\theta)$.
All curvature quantities discussed below therefore refer to this effective
two-dimensional geometry.

To determine the function \( \Omega(r) \), we consider the Einstein field equations in \( (2+1) \)-dimensional gravity~\cite{katanaev1992}:
\begin{equation}
    R_{\mu\nu} - \frac{1}{2} g_{\mu\nu} R = 8\pi G\, T_{\mu\nu},
    \label{eq:einstein_lambda}
\end{equation}
whose trace, in the absence of a cosmological constant, yields the scalar curvature in terms of the trace of the energy--momentum tensor:
\begin{equation}
    R = 8\pi G\,T ,
    \label{eq:scalar_curvature_trace1}
\end{equation}
where
\begin{equation}
    T = T^\mu_{\ \mu}.
\end{equation}

For the conformal metric~\eqref{eq:metric_conformal}, the scalar curvature is given by
\begin{equation}
    R =
    -2e^{-2\Omega}
    \left(
        \frac{d^{2}\Omega}{dr^{2}}
        +
        \frac{1}{r}\frac{d\Omega}{dr}
    \right).
    \label{eq:scalar_curvature}
\end{equation}

Substituting Eq.~\eqref{eq:scalar_curvature}
into Eq.~\eqref{eq:scalar_curvature_trace1}
yields the field equation for the conformal factor:
\begin{equation}
    \nabla^2 \Omega = -\lambda(r),
    \label{eq:omega_radial_lambda}
\end{equation}
where \( \lambda(r) \) represents the effective curvature density associated with the defect. To preserve the geometric interpretation of the source, this density must be defined with respect to the flat background, such that
\begin{equation}
    \lambda(r) = 4\pi G\, e^{2\Omega} T(r).
\end{equation}

This conformal formulation is well suited for modeling both singular and smooth distributions of disclinations, including delta-like cores, Gaussian profiles\cite{CarvalhoGarciaFurtado2026}, and multipolar arrangements~\cite{Carvalho2025}. It has been applied in studies of elastic media, polycrystalline textures, and synthetic materials where periodic curvature patterns emerge from structured defect networks~\cite{Ma2014,LuicanMayer2016}.

For an isolated disclination localized at the origin, the curvature density is modeled by a delta-function:
\begin{equation}
    \lambda(r) = \frac{1 - \alpha}{\alpha} \delta^{(2)}(\vec{r}),
    \label{eq:defect_density}
\end{equation}
where the parameter \( \alpha \) controls the angular defect. 
The curvature density given in Eq.~\ref{eq:defect_density} 
leads to the classical conical metric:
\begin{equation}
    ds^2 = dr^2 + \alpha^2 r^2 d\theta^2 + dz^2,
    \label{eq:conical_metric}
\end{equation}
This geometry has axial symmetry and a deficit angle \( \delta = 2\pi(1-\alpha) \).
The azimuthal angle \( \theta \) varies over the interval \( 0 \leq \theta \leq 2\pi\alpha \), so that for \( \alpha > 1 \) the defect corresponds to an angular excess, while for \( 0 < \alpha < 1 \) it corresponds to an angular deficit.
This convention is equivalent to the frequently used parametrization
\( \theta \in [0,2\pi) \) with metric component
\( g_{\theta\theta}=\alpha^2 r^2 \), both descriptions yielding the same
deficit angle \( \delta = 2\pi(1-\alpha) \).

It is locally flat for \( r > 0 \), with scalar curvature vanishing everywhere except at the origin.
The global structure, however, is nontrivial: parallel transport of vectors around the core produces a net rotation, evidencing a holonomy that reflects the topological nature of the defect~\cite{Furtado2008}.
Such a singular structure is characteristic of disclinations in elastic media and also appears in models of gravity in \( (2+1) \) dimensions~\cite{DeserJackiw1989}.

\section{Motivation for Including a Cosmological Term}
\label{motivation}

The inclusion of a cosmological constant \( \Lambda \) in the geometric description of topological defects is motivated by insights from analog gravity in topological media such as Weyl semimetals, superfluid \( ^3\text{He-A} \), and Bose–Einstein condensates~\cite{JannesVolovik2012,Finazzi2012,ORaifeartaigh2018cosmological}. In these systems, curvature and geometric responses arise effectively from deviations in the equilibrium structure of the vacuum, rather than from fundamental spacetime dynamics. The cosmological term can thus be interpreted as an emergent, scale-dependent correction that accounts for the system's nontrivial geometric response to topological or structural perturbations.

In this analogy, disclinations correspond to localized disruptions in rotational symmetry that deform the effective geometry experienced by quasiparticles. These regions may be seen as loci of partial relaxation, where the system does not fully return to its homogeneous ground state. As a result, a nonzero \( \Lambda \) emerges as a phenomenological parameter encoding the long-range modification of curvature induced by such defects.

From a geometric perspective, the cosmological constant modifies the Einstein field equations by introducing a uniform curvature term~\cite{Ellis2018,Carroll2001}:
\begin{equation}
    R_{\mu\nu} - \frac{1}{2} g_{\mu\nu} R + \Lambda g_{\mu\nu} = 8\pi G\, T_{\mu\nu},
    \label{eq:einstein_lambda2}
\end{equation}
Within the effective conformal framework adopted in this work, the presence
of a cosmological constant is incorporated through the modified
scalar-curvature relation
\begin{equation}
    R = 3\Lambda + 8\pi G\,T .
    \label{eq:scalar_curvature_trace_lambda}
\end{equation}

The additional term \( \propto \Lambda e^{2\Omega} \) renders the equation nonlinear and places it in the class of Liouville-type equations. This nonlinear contribution plays a key physical role: it sets the curvature scale of the geometry, controls its asymptotic structure, and leads to a smooth regularization of otherwise singular defects. Notably, similar Liouville-type structures arise in the dimensional reduction of Einstein gravity, where a cosmological constant term survives as a source in the effective two-dimensional theory~\cite{Grumiller2007}. In our case, this term encodes the response of the geometry to large-scale modulation and sets an infrared scale for curvature decay.

In the following section, we show that this modification leads to an exact, closed-form solution for the conformal factor in the presence of a delta-function disclination source. The resulting geometry interpolates between a classical conical core and an infrared-regular curved background, providing a scale-dependent generalization of the standard disclination model.

\section{Regularization via Cosmological Constant}
\label{sec:regularization}
We now derive the nonlinear Poisson equation from the trace relation introduced
in Sec.~4.   For the conformal metric, the scalar curvature is given by
\begin{equation}
R=-2e^{-2\Omega}\nabla^2\Omega .
\label{eq:scalar_from_laplacian}
\end{equation}
Substituting Eq.~\eqref{eq:scalar_from_laplacian} into
Eq.~\eqref{eq:scalar_curvature_trace_lambda} and multiplying both sides by
$-\frac{1}{2}e^{2\Omega}$, we obtain
\begin{equation}
\nabla^2\Omega
=
-\frac{3}{2}\Lambda e^{2\Omega}
-4\pi G e^{2\Omega}T .
\label{eq:nonlinear_poisson}
\end{equation}

For a point disclination located at the origin, the source term is modeled by
\begin{equation}
T=
\frac{1-\alpha}{4\pi G\,\alpha}\,
e^{-2\Omega}\delta^{(2)}(r).
\label{eq:source_disclination}
\end{equation}
Substituting Eq.~\eqref{eq:source_disclination} into
Eq.~\eqref{eq:nonlinear_poisson}, the conformal factors cancel exactly, yielding
\begin{equation}
\nabla^2\Omega
=
-\frac{1-\alpha}{\alpha}\,\delta^{(2)}(r)
-\frac{3}{2}\Lambda e^{2\Omega}.
\label{eq:poisson_alpha}
\end{equation}
Equation~\eqref{eq:poisson_alpha}
is a Liouville-type equation describing the interplay between a
localized topological defect and a cosmological-constant contribution
governed by $\Lambda$.
Assuming radial symmetry, its exact solution for $\Lambda>0$ is
\begin{equation}
\Omega(r)=
\ln\!\left[
\frac{2\sqrt{2}\,\alpha a}{\sqrt{3\Lambda}}
\frac{r^{\alpha-1}}
     {1+a^{2}r^{2\alpha}}
\right].
\label{eq:omega_solution_alpha}
\end{equation}
This closed-form expression provides the exact conformal factor for $\Lambda > 0$.
In this regularized model, the parameter $\alpha$ no longer acts as a global rescaling
of the angular sector but retains its role as the local conical strength near the defect.
In the limit $r \to 0$, the conformal factor is dominated by its logarithmic contribution,
and the solution asymptotically recovers the classical disclination geometry,
characterized by the angular deficit
$\delta = 2\pi(1 - \alpha)$.

The integration constant $a$ introduced in Eq.~\eqref{eq:omega_solution_alpha}
plays a central geometric role: it fixes the inverse length scale of the core, defining a characteristic crossover radius $r_c \sim a^{-1/\alpha}$
that separates the defect-dominated region from the infrared regime governed by the
cosmological term. While the near-core structure remains determined solely by $\alpha$, the cosmological constant $\Lambda$ sets the overall curvature scale of the geometry,
ensuring a smooth regularization of the defect without the need for artificial cutoffs.

It is important to emphasize that the closed-form expression for $\Omega(r)$ derived above is only valid for a positive cosmological constant ($\Lambda>0$), where the factor $\sqrt{3\Lambda}$ ensures a real, smooth, and physically meaningful conformal profile.
 This guarantees a regularized curvature and a constant-curvature geometry with
scalar curvature \(R=3\Lambda\).
For $\Lambda<0$, no real closed-form solution was found within the present
approach, and the corresponding profiles were therefore obtained numerically.
In this case, the geometry becomes asymptotically hyperbolic, with the scalar
curvature approaching the constant value $R \to 3\Lambda < 0$ at large
distances.

This qualitative difference between the signs of the cosmological term is consistent with reduced-gravity analyses such as Grumiller and Jackiw~\cite{Grumiller2007} and with earlier studies of Liouville gravity in lower-dimensional settings~\cite{Jackiw1985LowerDimensionalGravity,DeserJackiw1984LiouvilleGravity}, where exact solutions occur only for positive exponential coupling, while the negative case lacks closed-form metrics.

\begin{figure}[t]
  \centering
  \includegraphics[width=0.95\columnwidth]{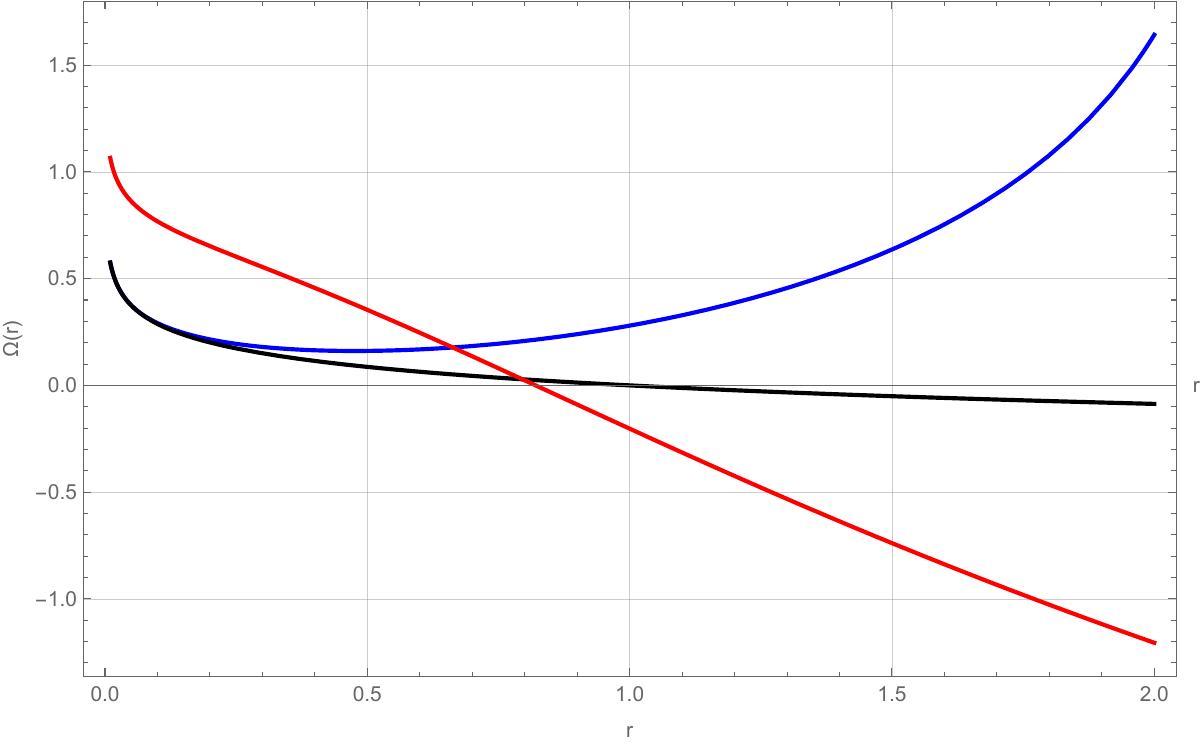}
\caption{
Radial profile of the conformal factor $\Omega(r)$ for a single disclination
with $\alpha=0.8$ and $a=1.0$, over $r\in[10^{-2},2]$.
Red curve: $\Lambda=+0.5$ (analytic solution);
black curve: $\Lambda=0$ (analytic);
blue curve: $\Lambda=-0.5$ (numerical solution).
All curves exhibit the same near-core logarithmic behavior,
$\Omega(r)\sim(\alpha-1)\ln r$,
showing that the local geometry remains governed by the disclination strength.
The cosmological constant modifies the large-distance behavior of the conformal factor,
leading to distinct infrared regimes for positive and negative values of $\Lambda$.
}
  \label{fig:omega}
\end{figure}

The behavior shown in Fig.~\ref{fig:omega} reveals that, close to the defect core, the behavior of $\Omega(r)$ is governed by the local conical structure and is essentially independent of $\Lambda$.
For $\Lambda>0$ (red curve), the Liouville term induces a stronger infrared decay of the conformal factor,
resulting in a regularized infrared geometry.
For $\Lambda<0$ (blue curve), $\Omega(r)$ bends upward beyond the crossover scale $r_c\sim a^{-1/\alpha}$,
signaling the transition to a hyperbolic-like geometry with negative curvature.
The $\Lambda=0$ case (black curve) interpolates between these regimes, corresponding to the classical conical geometry without large-distance modulation.

\subsection{Asymptotic Behavior and the Effective Geometry}

The conformal factor obtained from Eq.~\eqref{eq:omega_solution_alpha}
smoothly interpolates between a conical geometry in the vicinity of the core and a cosmological regime governed by the Liouville term at large distances.
The parameters $\alpha$, $a$, and $\Lambda$ determine, respectively, the angular deficit, the characteristic core size, and the curvature scale of the geometry. In the near-core limit $r\to0$, one has
\rev{$(1+a^2r^{2\alpha})^2\approx1$},
so the conformal factor reduces to
\begin{equation}
e^{2\Omega(r)}
\sim
\frac{8\alpha^2a^2}{3\Lambda}\,
r^{2\alpha-2},
\end{equation}

which yields the line element

\begin{equation}
ds^2
\approx
\frac{8\alpha^2a^2}{3\Lambda}
\left(
r^{2\alpha-2}dr^2
+
r^{2\alpha}d\theta^2
\right),
\label{eq:near_core_metric}
\end{equation}

recovering the angular deficit (or excess) characteristic of classical disclinations.

In the asymptotic regime $r\to\infty$, the denominator dominates as $(1+a^2r^{2\alpha})^2\approx a^4r^{4\alpha}$,
and the conformal factor behaves as

\begin{equation}
e^{2\Omega(r)}
\sim
\frac{8\alpha^2}{3\Lambda a^2}\,
r^{-2\alpha-2},
\end{equation}

so that the metric becomes

\begin{equation}
ds^2
\approx
\frac{8\alpha^2}{3\Lambda a^2}
\left(
r^{-2\alpha-2}dr^2
+
r^{-2\alpha}d\theta^2
\right).
\label{eq:asymptotic_metric}
\end{equation}

Both radial and angular parts decay rapidly, while the geometry approaches a constant-curvature regime characterized by \(R=3\Lambda\).
The cosmological constant $\Lambda$ sets the curvature scale of the asymptotic geometry, while the crossover scale $r_c\sim a^{-1/\alpha}$ marks the transition between the defect-dominated and cosmological regimes.

 This result shows that the scalar curvature is independent of both the disclination strength $\alpha$ and the core-scale parameter $a$. Substituting Eq.~(\ref{eq:omega_solution_alpha}) into Eq.~(\ref{eq:scalar_curvature}) yields $R=3\Lambda$ for all $r>0$. Therefore, the geometry possesses constant scalar curvature entirely determined by the cosmological constant. While $\alpha$ controls the local conical structure of the defect and $a$ sets the crossover scale of the conformal factor, neither parameter affects the curvature itself. This property reflects the Liouville character of the solution, in which the cosmological term uniquely determines the curvature of the background geometry. 
Although the scalar curvature is entirely determined by $\Lambda$, the disclination itself is not removed by the regularization mechanism. Its presence remains encoded in the nontrivial holonomy associated with the defect and in the conical geometry recovered in the near-core limit. Thus, the cosmological constant modifies the curvature background without erasing the topological charge carried by the disclination.

Introducing the radial coordinate
\begin{equation}
    \rho =
    \sqrt{\frac{8a^2}{3\Lambda}}\, r^\alpha ,
\end{equation}
one obtains
\begin{equation}
    d\rho =
    \sqrt{\frac{8a^2}{3\Lambda}}\,
    \alpha r^{\alpha-1}dr .
\end{equation}
Therefore, the near-core metric reduces exactly to
\begin{equation}
    ds^2 = d\rho^2+\alpha^2\rho^2d\theta^2 ,
\end{equation}
which is the standard conical metric of a classical disclination.

\section{Scalar Curvature Analysis}\label{sec:scalar}

To analyze the curvature of the regularized geometry, we evaluate the scalar curvature \(R(r)\) for a conformal metric \(g_{ij}=e^{2\Omega(r)}\delta_{ij}\). In two dimensions,
\begin{equation}
  R = -2\,e^{-2\Omega}\,\nabla^2\Omega .
\end{equation}

For the $\Lambda>0$ branch, the radial Laplacian of the solution $\Omega(r)$ yields
\begin{equation}
  \nabla^2 \Omega =
  -\frac{4\alpha^2 a^2 r^{2\alpha-2}}
  {\bigl(1+a^2 r^{2\alpha}\bigr)^2},
  \label{eq:laplacian_omega}
\end{equation}
and the conformal factor is
\begin{equation}
  e^{2\Omega(r)}
  =
  \frac{8\alpha^2a^2}{3\Lambda}\,
  \frac{r^{2\alpha-2}}
  {\bigl(1+a^2 r^{2\alpha}\bigr)^2}.
  \label{eq:conformal_factor_scalar}
\end{equation}
Hence
\begin{equation}
  e^{-2\Omega(r)}
  =
  \frac{3\Lambda}{8\alpha^2a^2}\,
  r^{2-2\alpha}
  \bigl(1+a^2 r^{2\alpha}\bigr)^2,
\end{equation}
and substituting into Eq.~\eqref{eq:scalar_from_laplacian} gives the closed-form scalar curvature, valid for all $r>0$: $ R = 3\Lambda$

The cancellation of the integration constant $a$ confirms that the curvature is independent of the crossover scale ($r_c\sim a^{-1/\alpha}$).  The defect parameters $\alpha$ and $a$ affect the local geometry, but not the scalar curvature itself, which remains constant throughout the regularized space.
Therefore, the cosmological constant provides a natural geometric regularization of the disclination without altering its topological character. The resulting solution preserves the local conical structure near the core while embedding the defect into a smooth constant-curvature background. This interplay between topology and curvature distinguishes the present construction from the singular conical geometry and motivates a more detailed analysis of the global geometric properties.


\subsection{Negative Cosmological Constant}
For $\Lambda<0$, no real closed-form solution was found within the present approach, and the corresponding geometries were therefore investigated numerically. The numerical solutions indicate that the geometry evolves from the conical core toward an asymptotically hyperbolic regime, with the scalar curvature approaching
\begin{equation}
R \longrightarrow 3\Lambda <0.
\end{equation}
This behavior contrasts with the regularized positive-curvature geometry obtained for $\Lambda>0$ and highlights the role of the cosmological constant in determining the asymptotic structure of the defect geometry.


\section{Conclusions}
\label{conclusion}

In this work, we investigated the impact of a nonzero cosmological constant on the geometry generated by a disclination in $(2+1)$-dimensional space using a conformal metric framework. For positive cosmological constant, we obtained an exact analytic solution to the nonlinear Liouville-type equation with a point-like source, which regularizes the defect geometry while preserving its local conical structure. For negative cosmological constant, no real closed-form solution was found within the present approach, and the corresponding geometries were therefore investigated numerically.
The results reveal a clear geometric dichotomy. For $\Lambda>0$, the exact solution describes a regularized constant-curvature geometry with $R=3\Lambda$, while preserving the local conical structure of the defect. For $\Lambda<0$, the numerical solutions approach an asymptotically hyperbolic geometry with $R \rightarrow 3\Lambda < 0$. The sign of $\Lambda$ therefore determines the asymptotic geometry of the regularized defect, whereas the disclination strength $\alpha$ remains encoded in its local conical topology.

The conformal approach adopted here offers a transparent and unified description of disclinations in the presence of a cosmological constant, without resorting to arbitrary cutoffs. It naturally connects to effective geometries in analog gravity systems, such as superfluids and Bose--Einstein condensates, where emergent vacuum terms mimic cosmological behavior. Future investigations may explore geodesics, holonomy, quantum phases, and the extension to more complex defect networks, possibly including torsion. Overall, the cosmological constant emerges not only as a natural geometric regularization mechanism but also as the parameter that determines the curvature scale and global structure of the defect geometry. The resulting interplay between topology and constant-curvature backgrounds provides a simple framework for investigating regularized disclinations and related analog-gravity systems, with potential applications to realistic condensed matter models such as graphene~\cite{vozmediano2010gauge,cortijo2007effects}.

\section*{Data Availability}
No data are associated with this manuscript

\section*{Disclaimer} ChatGPT was used only as an auxiliary editorial tool for limited language refinement, text organization, formatting suggestions, and readability improvements. It also provided occasional assistance in checking consistency and clarity of presentation. All scientific ideas, research design, literature analysis, mathematical derivations, physical interpretations, results, conclusions, and editorial decisions were developed and validated exclusively by the authors, who assume full responsibility for the content of this manuscript.

{\bf Acknowledgements:}  The work by C. Furtado is supported by the CNPq (project PQ Grant 1A No. 311781/2021-7).



\end{document}